\documentclass{aa}

\usepackage{natbib}
\usepackage{graphicx}
\newenvironment{jh}{}{}

\newcommand{\micron}{\hbox{$\mu{\rm m}$}}                      

\newcommand{\msun}{\mbox{$M_{\odot}$}}
\newcommand{\Msun}{\mbox{$M_{\odot}$}}

\newcommand{\Tdust}{\mbox{$T_\mathrm{d}$}}

\newcommand{\Lsun}{\mbox{$L_{\odot}$}}

\defcitealias{paperI}{Paper~I}
\defcitealias{class}{Paper~II}
\defcitealias{outflows}{Paper~III}

\begin{document}
   \title{Star formation in Perseus}

   \subtitle{IV. Mass dependent evolution of dense cores.}

   \author{J. Hatchell\inst{1}, G. A. Fuller\inst{2}}
   \authorrunning{Hatchell et al.}
   \titlerunning{SCUBA Perseus survey -- IV. Mass dependent evolution}
   
   \offprints{hatchell@astro.ex.ac.uk}

   \institute{School of Physics, University of Exeter, Stocker Road, Exeter EX4 4QL, U.K.
              \and Jodrell Bank Centre for Astrophysics, Alan Turing Building,
             University of Manchester,  Manchester M13 9PL, U.K.
              }

   \date{Feb 08}
   
   \abstract{In our SCUBA survey of Perseus, we find that the fraction of
     protostellar cores increases towards higher masses and the most massive
     cores are all protostellar.}
   {In this paper we consider the possible explanations of this
       apparent mass dependence in the evolutionary status of these
       cores. We investigate the implications for protostellar
     evolution and the mapping of the embedded core mass function
     (CMF) onto the stellar IMF.}
   {We consider the following potential origins of the observed
       behaviour: dust temperature;  selection effects in the submillimetre and in the mid-infrared observations
     used for pre/protostellar classification; confusion and multiplicity;
     transient cores; and varying evolutionary timescales.  We develop Core Mass
     Evolution Diagrams (CMEDs) to investigate how the mass evolution
     of individual cores maps onto the observed CMF.}
   {We find that two physical mechanisms -- short timescales for
       the evolution of  massive cores, and continuing accumulation of mass \begin{jh}onto protostellar cores\end{jh} -- best explain the  relative excess of
       protostars in high mass cores and the rarity of massive
       starless cores.  In addition, we show that confusion
      both increases the likelihood that a protostar is
     identified within a core, and increases mass assigned to a
       core.  Selection effects and/or transient cores also
     contribute to an excess of starless cores at low masses.  }
   {The observed pre/protostellar mass distributions are consistent with faster evolution and a shorter lifetime for higher-mass prestellar cores.  We rule out longer timescales for higher-mass prestellar cores. 
  The differences in the prestellar and protostellar mass
     distributions imply that the prestellar CMF (and possibly the
     combined pre+protostellar CMF) should be
steeper 
than the IMF.  A steeper prestellar CMF can be reconciled with the
observed similarity of the CMF and the IMF in some regions if a
second opposing effect is present, such as the fragmentation of
massive cores into multiple systems. 
}

\keywords{Submillimeter;Stars: formation;ISM: clouds}

\maketitle

\section{Introduction}
\label{sect:introduction}

The stellar Initial Mass Function (IMF) {\em may} 
largely be determined in the prestellar phase of the evolution of
dense cores, as the close agreement between the form of the prestellar
core mass function (CMF) and the IMF enticingly suggests
\citep{motte98,testisargent98,johnstone00,johnstone01,motte01,johnstone06,reidwilson06,kirk06,nutter07,andre07,alves07,goodwin07}.
However, feedback, competitive accretion and varying evolutionary
timescales may all have a role to play in the 
evolution of the cores.  Clearly, it is important to study how the
mass in dense starless cores relates to the mass in cores forming
protostars (protostellar cores), and therefore what determines the
mass available to each protostar as it grows in mass.

There are now a number of molecular clouds for which the dense cores
have been catalogued (from millimetre (mm) and submillimetre (submm)
surveys) and classified (using the mid-infrared detections from the
Spitzer Space Telescope).  Perseus is one such cloud, for which we
carried out a submm survey with SCUBA
(\citealt{paperI,class,outflows}, hereafter Papers~I--III, see also
\citealt{kirk06} and the similar 1300\micron\ survey with Bolocam by
\citealp{enoch06}).  We have identified over 100 submm cores, each of
which has been classified as starless or protostellar using Spitzer
(\citetalias{class},\citealt{jorgensen06a,jorgensen07}) and molecular outflows
\citepalias{outflows}.

An intriguing result in Perseus is that the highest mass cores are all
protostellar (Fig.~\ref{fig:masses}), and as one moves towards lower
masses, an increasing fraction of cores appear to be starless.  
This leads to the obvious questions: 
why are there so few massive starless cores and why are the two
distributions different?  In this paper we address these
  questions.  We start by first reviewing the results of the previous
work on this cloud (Sect.~\ref{sect:results}), and consider the
  effects of the analysis of the submm data, detection limits and
multiplicity on the distributions.  Then in Sect.~\ref{sect:evolution}
we discuss some evolutionary scenarios which might produce these
effects \begin{jh}and introduce core mass evolution diagrams to illustrate how core evolution maps onto the observed CMF.  In Section~\ref{sect:discussion} we explore the factors which affect the relationship between the CMF and the IMF before presenting the summary and conclusions in Section~\ref{sect:summary}.\end{jh}

\section{Results}
\label{sect:results}


The 103 submm cores discussed here were identified using Clumpfind
\citep{williams94} on the SCUBA 850\micron\ map of Perseus spatially
prefiltered by removal of a $2'$ Gaussian smoothed background
(primarily to remove artefacts from the image reconstruction; see
\citetalias{class} for details).  The identified cores have
masses in the range 0.5--50~\msun, sizes $\sim 0.1$~pc ($1'$ at the
assumed 320~pc distance of Perseus), and mean H$_2$ densities of
$>3\times 10^5\hbox{ cm}^{-3}$.

In \citetalias{class}, we classified the cores using mid-IR detections
from Spitzer to identify protostars.  There are slight discrepancies between our classification and the
embedded YSO list of \citet{jorgensen07}, due to the different
classification methods used (in our case, IRAC sources within $12''$ or
the Clumpfind source radius of the submm peak
\citepalias[see][]{class}, whereas \citet{jorgensen07} looked for
MIPS~24\micron\ sources within $15''$ of the peak).  In total, 13 of
our sources are classified differently by \citet{jorgensen07}, but \begin{jh}only one of our massive protostellar sources ($M > 1~\Msun$) is reclassified as starless.\end{jh}  The majority of the
massive protostellar sources are well-known 
(L1448C, L1448~NW, NGC1333~IRAS~4A, HH211).  Therefore the
alternative classification does not significantly change the
distributions.  We found
that the ratio of protostellar cores to starless cores increased with
mass, with ultimately no starless cores at all at the highest masses
(above 12~\msun, \citetalias[][ Fig.~3]{paperI}).
This result is shown in a modified form in \begin{jh}the top panel of \end{jh}Figure~\ref{fig:masses}(Section~\ref{sect:temperature}).

In \citetalias{outflows} we used an alternative classification \begin{jh}of the
nature of the cores \end{jh}based on the presence or absence of molecular
outflows for a subsample of 51 sources, and found the same result.

In fact these effects (the most massive cores are
protostellar, with more starless cores at lower masses) have now been
independently confirmed in Perseus by \citet{enoch08} using masses based on
Bolocam~1.3\micron\ data \citep{enoch06}, and the alternative Spitzer
identification \citep{jorgensen06a,rebull07,jorgensen07,lai07}.  \citet{enoch08} also finds the same
holds for Serpens, though intriguingly not Ophiuchus, which shows more
similar mass distributions for starless and protostellar cores. The
massive star-forming region Cygnus~X shows a more extreme version of
this effect, with no massive starless cores \citep{motte07}.

\section{Assumptions and selection effects}
\label{sect:assumptions}

Below we consider possible selection effects and assumptions which could
influence the our results and the core mass distributions.

\subsection{Detection and Identification}
\label{sect:detection}


The submm continuum detections are flux-limited.  The observations are
not sensitive to cores below a column density threshold of $8\times
10^{22} \hbox{ cm}^{-2} (A_v \simeq 90)$ (the $5\sigma$ flux detection
limit at 850\micron).  For a point source, this corresponds to a mass
limit of $0.5$~\Msun (assuming 10~K and opacity $\kappa_{850} =
0.012~\hbox{cm}^2\hbox{ g}^{-1}$, see \citetalias{class}), but more
massive sources can go undetected if the flux is more extended.  Below
this, our core detection is incomplete, and deeper studies are needed
to determine the form of the mass distribution \citep[eg. the Orion
metastudy by ][]{nutter07}.  The sensitivity and high anticipated
source counts of the next generation of submillimetre surveys, such as
the JCMT Gould's Belt Legacy Survey \citep{JCMTGB} and Herschel survey
of nearby star-forming regions \citep{andresaraceno05} will lay to
rest incompleteness issues at the 0.1~\msun\ level in nearby clouds.
But currently, with a relatively shallow column density detection
limit, for many cores we must be only detecting the `tip of the
iceberg' with much of the mass lying at lower, undetectable column
densities.  In addition, objects over $2'$ (0.2~pc) in size are
significantly affected by the spatial filtering of the data needed to
removed artefacts on large size scales.  The impact of the filtering
depends on the degree of central concentration of a source with the
less condensed sources being the most strongly suppressed.


For both these reasons (high column density detection threshold and
filtering of extended cores) we are better at detecting cores with a
high degree of central concentration, so the observations may only be
tracing the later stages of prestellar evolution
\citep[eg.][]{tassismouschovias04}.  For cores smaller than $\sim2'$
in size, higher mass cores are easier to detect.  A priori we might
expect a selection effect which favours the detection of starless (and
protostellar) cores with higher masses.
So the lack of high-mass starless cores is not a consequence of our inability
to detect them if they exist.  The submm detection limits do not explain the
lack of high-mass starless cores.

\subsection{Temperature}
\label{sect:temperature}

\begin{figure}[t]
\centering
\includegraphics[scale=0.90,angle=-90]{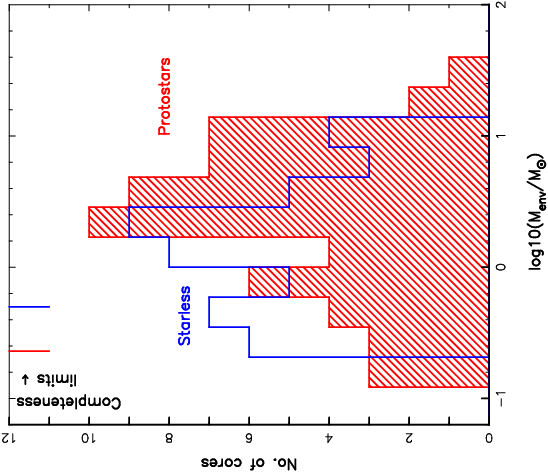}
\includegraphics[scale=0.90,angle=-90]{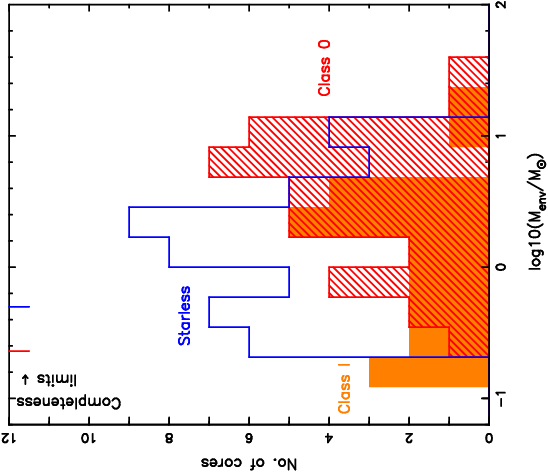}

\caption{{\bf Top:} Histogram of masses of submm cores, assuming 10~K for
  starless cores (blue line) and 15~K for protostars (red
  hatched).  The SCUBA completeness limits for point sources at each temperature are
  marked at the top of the plot. {\bf Bottom:} As above with protostars separated into Class~0 (red hatched) and Class~I (orange filled). }

\label{fig:masses}
\end{figure}

%
%

\begin{figure}[t]
\centering
\includegraphics[scale=1.0,angle=-90]{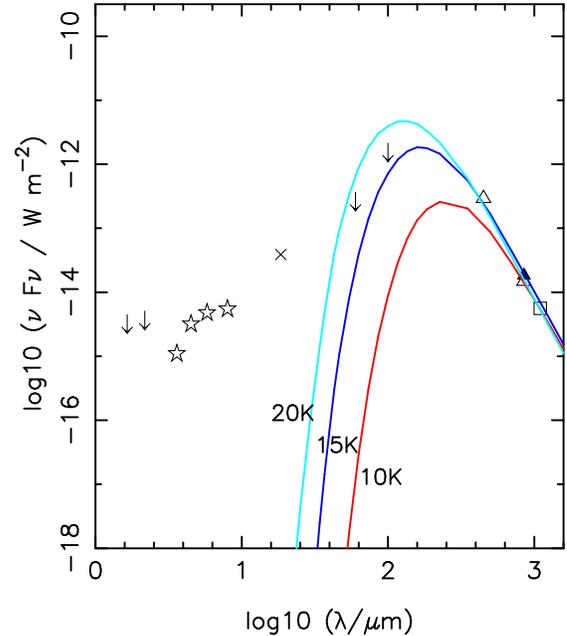}

\caption{Average SED for all 34 Class~0 sources overplotted with
  modified blackbodies at 10, 15 and 20~K.  Flux points are the mean
  of all detections from Class~0 sources at each wavelength, from
  (left to right) 2MASS, Spitzer IRAC, Michelle, IRAS, SCUBA and
  Bolocam \citepalias[see][for details]{class}.  The 70 and 100~micron
  points are strictly upper limits because of potential confusion
  within the IRAS beam, but still constrain the maximum temperature to
  less than 20~K.  The 2MASS datapoints are marked as upper limits as
  the high fluxes are likely due to confusing sources. }

\label{fig:aveclass0sed}
\end{figure}


In \citetalias{class}, the masses of the cores were calculated from
the 850\micron\ emission assuming a constant dust temperature of 10~K,
which was the highest estimate for \Tdust\ consistent with all the
spectral energy distributions (SEDs).  However, protostellar sources
tend to have higher temperatures and so produce relatively more
850\micron\ emission for the same mass of circumstellar material.  Can
the assumed dust temperature reconcile the masses of the starless and
protostellar cores?

In order to test this, we recalculate the masses for the sources
assuming different dust temperatures.   Recent studies of
  NH$_3$ found a small difference between {\em gas} temperatures of 11 and 12~K for starless and protostellar cores,
  respectively \citep{rosolowsky07}.  
However, differences in the dust temperature between starless and protostellar cores are evident from continuum observations of the SEDs of sources \citep{class,enoch08}.  The apparent difference between the gas and dust temperatures, averaged along the line of sight, is presumably due to the different sensitivity of molecular line and dust emission to the warmer inner regions of the sources.
Modelling of prestellar and protostellar cores has found equivalent
isothermal dust temperatures of $T_\mathrm{dust} = 11$~K (starless),
$14$~K (Class~0 \begin{jh}protostars\end{jh}) and 16~K (Class~I \begin{jh}protostars\end{jh}) \citep{evans01,shirley02,young03}.

We follow \citet{enoch06} in assuming 10~K for
starless cores and 15~K for protostars, in order to
  ensure that we account fully for dust temperature
  differences.  The mass distribution for our sources
assuming \begin{jh}10~K and 15~K\end{jh} temperatures for the starless and protostellar
sources respectively is shown in \begin{jh}the upper panel of \end{jh} Figure~\ref{fig:masses}.  Although
reduced from \citetalias{class}, due to the lower masses inferred for the protostellar
sources, there is still a significant difference in the mass
distribution of the starless and protostellar cores, with more
protostellar sources at higher masses and more starless cores at lower
masses.  To quantify this, in the whole population 46\% of sources are
starless, but above 3~\Msun\ this falls to 31\%, 20\% above 10~\Msun,
and zero above 15~\Msun.  A K-S indicates that the probability that
these two samples are drawn from the same mass distribution is 4\%,
compared to 0.001\% when a temperature of 10K is assumed for all the
sources as in \citetalias{class}.  Only considering sources above
1~\Msun, the K-S test gives a probability of 8\% that the
distributions are the same.

\begin{jh}The lower panel of Figure~\ref{fig:masses} shows the
  core mass distribution but with the protostars separated into Class~0 and Class~I
  evolutionary stages based on the classification in \citetalias{class}.
  As expected, the more evolved Class~I cores have lower masses than
  the Class~0 sources.  From this figure we see that
  the excess of protostars is mainly due to the excess of Class~0
  cores, and therefore it is the temperature of the Class~0 population
  that is critical to the difference between the prestellar and protostellar mass distributions.\end{jh} Since the sense in which
the mass distributions \begin{jh}differ\end{jh} is that the Class~0
sources are more massive, this difference could be reduced if the
Class~0 sources had even higher temperatures.  Raising the assumed
temperature to 20~K for the Class~0 sources would produce a
distribution more compatible with that of starless cores (15\% K-S
probability).  However, a temperature as high as 20~K (which would be
larger than the temperature of the Class~I sources) is ruled out by
the individual SEDs, including those of the most massive Class~0 cores
\citepalias{class}.  The average SED for the Class~0 cores is shown in
Fig.~\ref{fig:aveclass0sed}.  As the model SEDs show, the
observations are compatible with a temperature of 15~K but not 20~K.
Therefore even taking temperature into account, there is still a
significant difference between the two mass distributions.

A consequence of assuming a temperature of 15~K rather than 10~K is
that a given 850\micron\ flux corresponds to a factor 2 lower column
density.  As Clumpfind integrates mass down to a fixed flux limit, the
mass of higher temperature gas is measured down to a lower column
density than lower temperature gas.  If temperatures at the outer
edges of protostars are significantly higher than those for starless
cores, this creates a bias towards higher masses for protostars as
surrounding material at lower column densities is included than for
starless cores.  To quantify this effect requires an understanding the
temperature structure of the cores.  The 15~K and 10~K values are the
mass-weighted average in the beam, with the higher value for
protostars due to the central heating.  At large radii, the
temperatures of protostars and starless cores are more similar.
Models suggest that temperatures fall to below about 12~K within
10,000~AU ($30''$) of the central protostar and 
both
starless 
and protostellar cores are heated to 12-14~K at their outer edges by
the interstellar radiation field \citep{evans01,shirley02}, 
suggesting that this possible bias is not likely to produce a significant
  systematic effect.

In addition to the different mass distributions of the
  protostellar and starless cores, a slightly different question is
  whether there is a significant excess of protostars associated with
  higher mass cores. 
Above 10~\Msun, only 2 out of 10 sources are starless.
Applying the binomial distribution, the probability of getting 2 or
fewer starless cores by chance is 9.3\%.  Although not entirely
impossible that this is a random fluctuation (and we are limited here
by the small number of high mass objects in one cloud), the
overpopulation of protostars at high masses is likely to be real.
This is further supported by the excess of high mass protostars also
found in Serpens \citep{enoch08}.



\subsection{Source Confusion}
\label{sect:confusion}

%

The absence of massive starless cores could result from source confusion, if
our analysis erroneously fails to separate multiple lower mass cores, one of
which happens to contain a protostar.  This could occur due to insufficient
angular resolution, or overlapping cores along the line of sight. Both of
these are potentially more significant issues in regions where cores and
protostars are clustered, such as NGC1333 and IC348.

The 10 most massive protostellar cores in our sample are all in
clusters or groups (B1, IC348, L1448, NGC1333).  This is consistent
with the expectation that massive stars form in clusters.  We might
also expect to find the most massive prestellar cores in clusters.
Yet only two of the four most massive prestellar cores are in NGC1333;
the other two are the filaments to the SW of B1 and the SE of L1448,
outside the main clusters of protostars.  It is possible that we fail
to identify starless cores in clusters because they are confused with
existing protostars.  Many of the most massive cores contain multiple
protostars: L1448~NW, NGC1333~IRAS~4A and 4B, NGC1333~IRAS 2A/B.  The
core B1-b contains multiple submm peaks \citep{hirano99} but only one
object is identified as a protostar.  

In clusters, it is often unclear
how to share the mass in the dense, fragmenting filaments between
individual objects.  Multiplicity clearly affects our ability to
identify, at the prestellar and protostellar stage, the amount of mass
which ultimately will contribute to one star.  As protostars produce
strong peaks, and all clumpfinding software looks for peaks, some
prestellar material may be being attributed to nearby protostars,
increasing the mass of the protostellar cores at the expense of
starless cores.  Thus, confusion could explain why we identify the
most massive protostellar cores in clusters.  However, it does
not explain why only the more massive starless cores are underabundant
with respect to protostars, as we would expect at least as many
lower-mass starless cores to be lost due to confusion.


\subsection{Stellar Content}
\label{sect:stellarcontent}

The difference between the mass distributions also depends on our
ability to differentiate protostars from starless cores.  For
protostars, luminosity and outflow power decrease with decreasing mass
\citep{batc96,fullerladd02}, and therefore both the MIR and CO outflow
emission become weaker.   The mid-IR and outflow detection limits
could therefore explain why there are more apparently starless cores
at lower masses.

The Spitzer c2d survey \citep{c2d} is limited in the lowest luminosity
source it can detect.  The Spitzer detections for the lowest-mass
sources identified as protostars are all close to the IRAC detection
limits and the addition of MIPS-only detections increases the number
only by a few \citep{rebull07,jorgensen07}.  Spitzer c2d is estimated to be 100\%
complete to a luminosity of 0.3~\Lsun\ in Serpens \citep{harvey07}
though significant fractions of lower-luminosity sources are detected
\citep[e.g.][]{dunham07}, with 50\% completeness at 0.01~\Lsun.  We
can assume these limits also apply to Perseus, as it is at a similar
distance.  So the population of protostars in Perseus with
$L<0.3\Lsun$ remains incompletely sampled by Spitzer.  For reference,
a luminosity limit of 0.3~\Lsun\ implies completeness to sources above
1~\Msun\ final stellar mass, according to the PMS tracks of
\citet{siess99} (assuming constant accretion). Some of the low-mass
starless cores could contain undetected low-luminosity MIR sources and
this will only be resolved by further deep searches in the MIR, such
as with {\em Herschel} \citep{andresaraceno05}.

In \citetalias{outflows} we considered in detail the detection
statistics for outflows and concluded that the outflow observations
also do not conclusively rule out the presence of low mass, low
luminosity protostars in the apparently starless cores.  Subsequent
observations suggest that the majority of the cores are indeed
starless, i.e. they do not contain outflows, when a deeper (by factor
$> 5$) outflow search is carried out (Hatchell et al. in prep.).

Nonetheless, although incompleteness might explain an excess of
  starless cores, it can only increase the fraction of starless cores as
  compared to protostars. In particular the failure of MIR and
outflow observations to detect low-luminosity sources cannot explain
why all the most massive cores are protostellar.

\subsection{Transient cores}
\label{sect:transience}

A population of transient cores, cores which are not
  gravitationally bound, might contribute to explaining the
overpopulation of low-mass starless cores.  A population of failed
cores which will not form protostars or transient starless cores which
increases at lower masses was suggested by \citet{elmegreen97} and is
seen in turbulence simulations \citep{padoan02,vazquezsemadeni05b}.
Cores whose chemistry suggests transience are also observed
\citep{moratagirart05}.

  Estimates of the fraction of cloud mass in dense cores range from a few percent \citep{fuller&myers87,johnstone04,motte98} to a few tens of percent \citepalias{paperI}, typically with uncertainties of more than a factor of two.    Ultimately, the mass expected to end up in stars is also few percent of the total cloud mass \citep{ladalada03}, suggesting that in some cases there will be little mass in the core budget which does not ultimately go in to stars.

It has been argued that cores that are centrally condensed enough to
be detected in the submm must be forming stars and therefore are not
transient \citep{motteandre01}.  With high mean densities above
$n_{H_2} \sim 3\times 10^{5} \hbox{cm}^{-3}$, corresponding to a thermal Jeans mass
of 0.3~\msun), our sources contain from one to tens of Jeans masses
and thus satisfy the minimum condition to be gravitationally bound.
On the scales probed by the N$_2$H$^{+}$ line (critical density $10^5
\hbox{cm}^{-3}$), linewidths are roughly thermal and the majority of
cores are virially supported if an external pressure term is included
\citep{kirk07}.  It therefore seems unlikely that we are seeing a
large population of unbound objects.  However, it is
possible that some cores could be destroyed by future energetic
events, particularly cores in disruptive cluster regions, or that the currently binding external pressure decreases to the point where cores re-expand, particularly for lower-mass cores.

\section{Core mass evolution}
\label{sect:evolution}

\begin{figure*}[t]
\centering
\begin{tabular}{c c c }
\includegraphics[scale=0.3,angle=0]{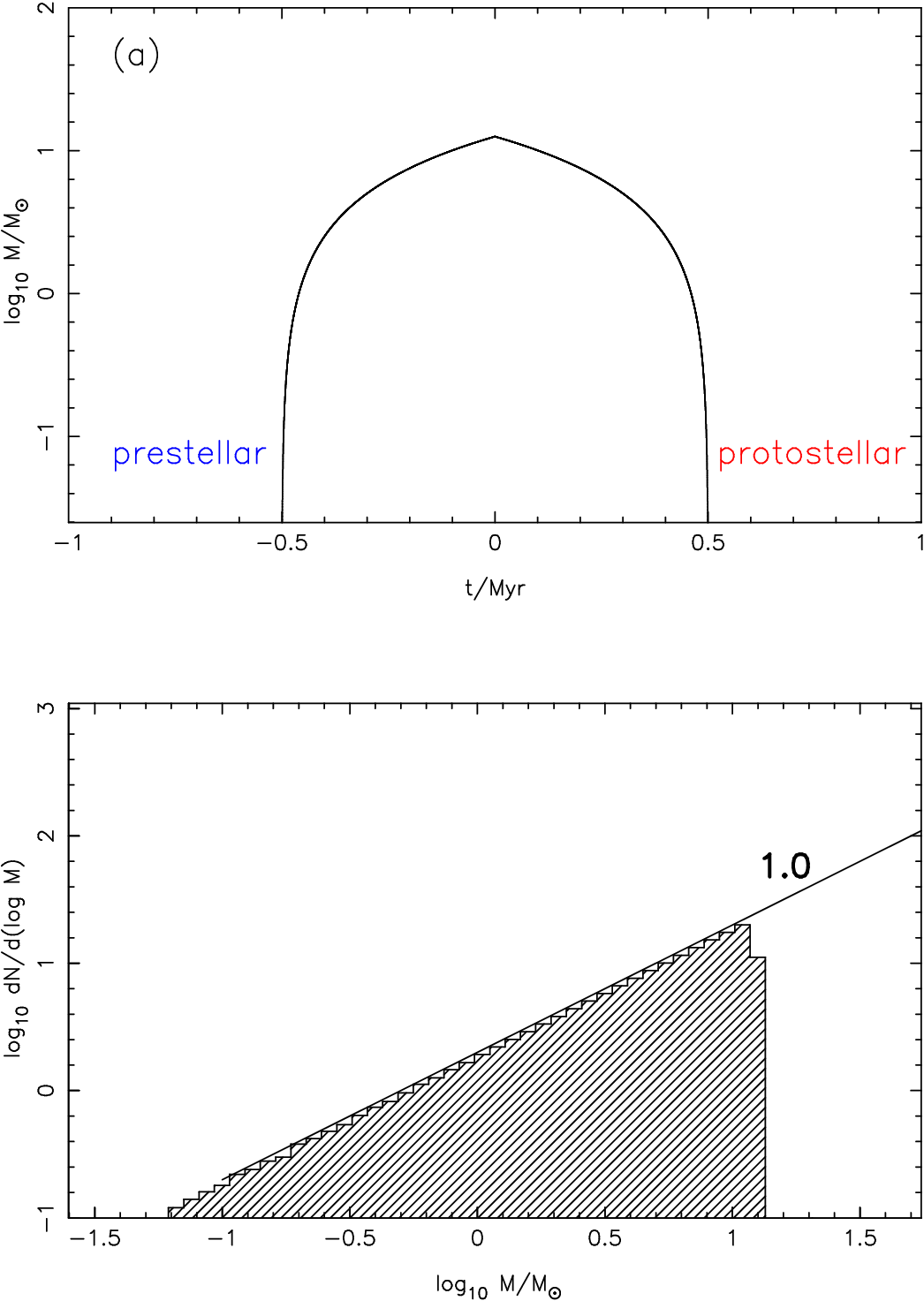}&
\includegraphics[scale=0.3,angle=0]{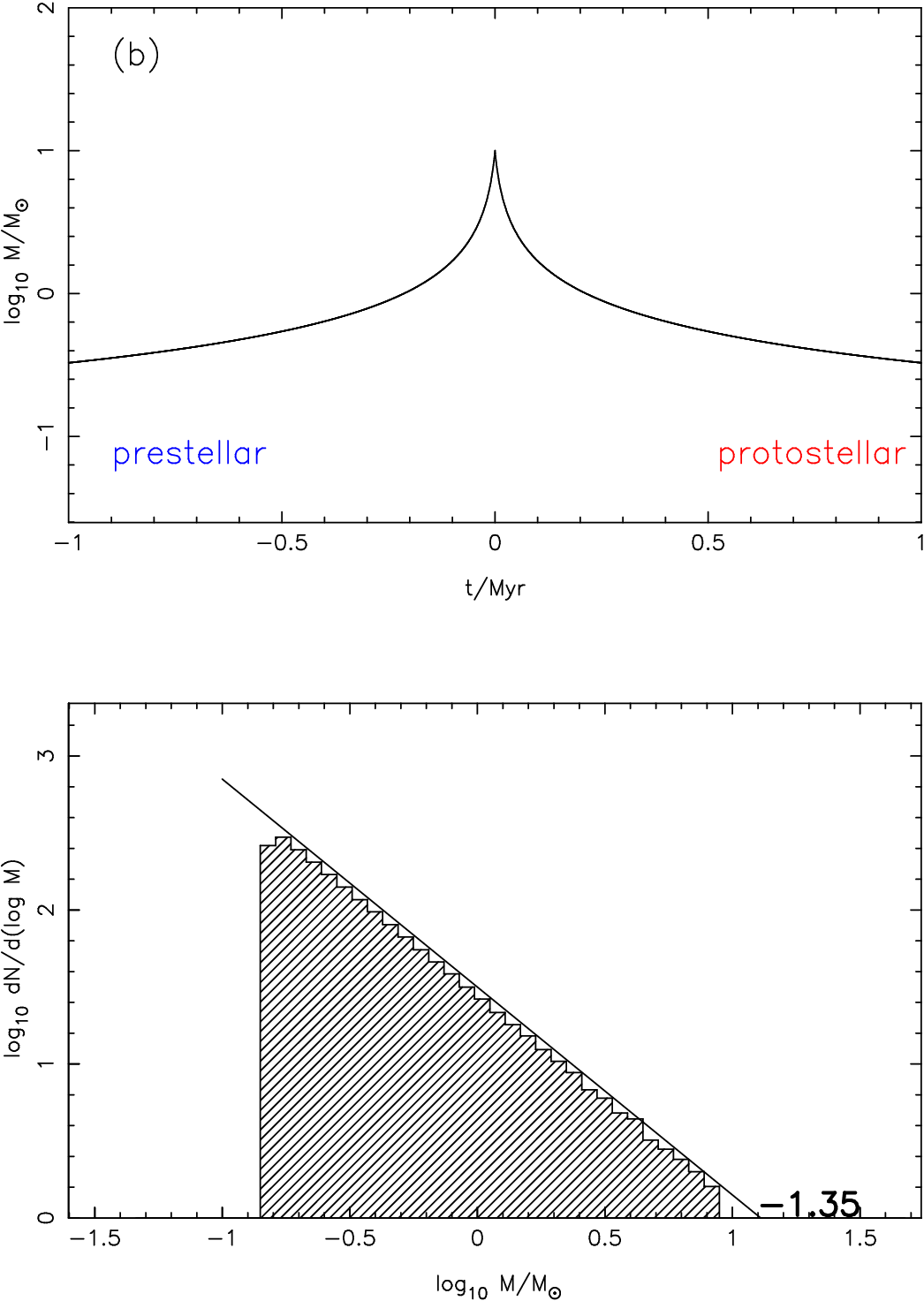}&
\end{tabular}
\caption{Core mass evolution diagrams (CMEDs) for single evolutionary tracks.  Each top panel shows on a linear-log plot the evolution of the core mass (as measured above a detection threshold) as a function of time.  The bottom panels show the corresponding core mass functions assuming a population of identical cores sampled at random times. {\bf (a):} Linear core mass evolution.  {\bf (b):} The pathological case where a population of identical cores results in a Salpeter power law CMF simply through mass evolution governed by Equation~\ref{eqn:salpeter}.}
\label{fig:sim_single}
\end{figure*}


The observed prestellar and protostellar core mass functions are built
up of mass measurements of many individual sources.  To fully
understand them we must consider how individual objects evolve and how
they appear to the observer at different times.  By modelling the mass
evolution of cores, and reflecting this in the CMF, we have a tool to
investigate how different factors in core evolution can lead to
differences in the observed CMFs.

One possibility is that the timescale for forming a star,
$t_\mathrm{sf}$, may be dependent on core mass.  As the number of
prestellar or protostellar objects is the product of the production rate
and the lifetime, longer lifetimes in a particular phase imply a
greater number of detections and a large representation on the CMF.

A second factor is 
the relationship between the time it takes a core to form a star, and
the time it takes for the core to grow in mass.  The important
parameter here is the ratio the timescale for a core of a given mass
to form a protostar ($t_{\rm sf}$) to the timescale for the core mass
to grow through the accretion of additional material from its
surrounding cloud, ($t_{\rm gr}$).  In the limit $t_{\rm sf}\ll t_{\rm
  gr}$, a core can be considered as being fixed mass reservoir while
it is forming stars.

On the other hand, if $t_{\rm sf}\stackrel{>}{_\sim} t_{\rm gr}$, star
formation will be taking place within a core as the core continues to
grow in mass. These two scenarios have different implications for the
relationship between the mass distributions of starless and
protostellar cores.




\subsection{Core mass evolution and the CMF}
\label{sect:CMED}

The detectable mass of starless cores and protostars changes as they
evolve, so that each object moves across a range in observable masses
during its lifetime, and the CMF is an average of sources at different
stages in their evolution.  Prestellar cores start as diffuse objects
below the column density detection limits of submm surveys such as
this one (see Sect.~\ref{sect:detection}), and become more centrally
condensed with time so that the detectable mass starts at the
detection threshold and rises throughout the prestellar phase.  For
example, in the seminal collapse calculation of \citet{larson69},
detectable central densities of $3\times 10^5$~cm$^{-3}$ are reached after
$2\times 10^5$~yr and the mass above the detection threshold continues
to rise until a central core is formed at $5\times 10^5$~yr.  For
protostellar cores, outflows and accretion remove mass, and the
Class~I sources have lower masses than Class~0s
(Fig.~\ref{fig:masses}).

\begin{figure*}[t]
\centering
\begin{tabular}{c c c }
\includegraphics[scale=0.3,angle=0]{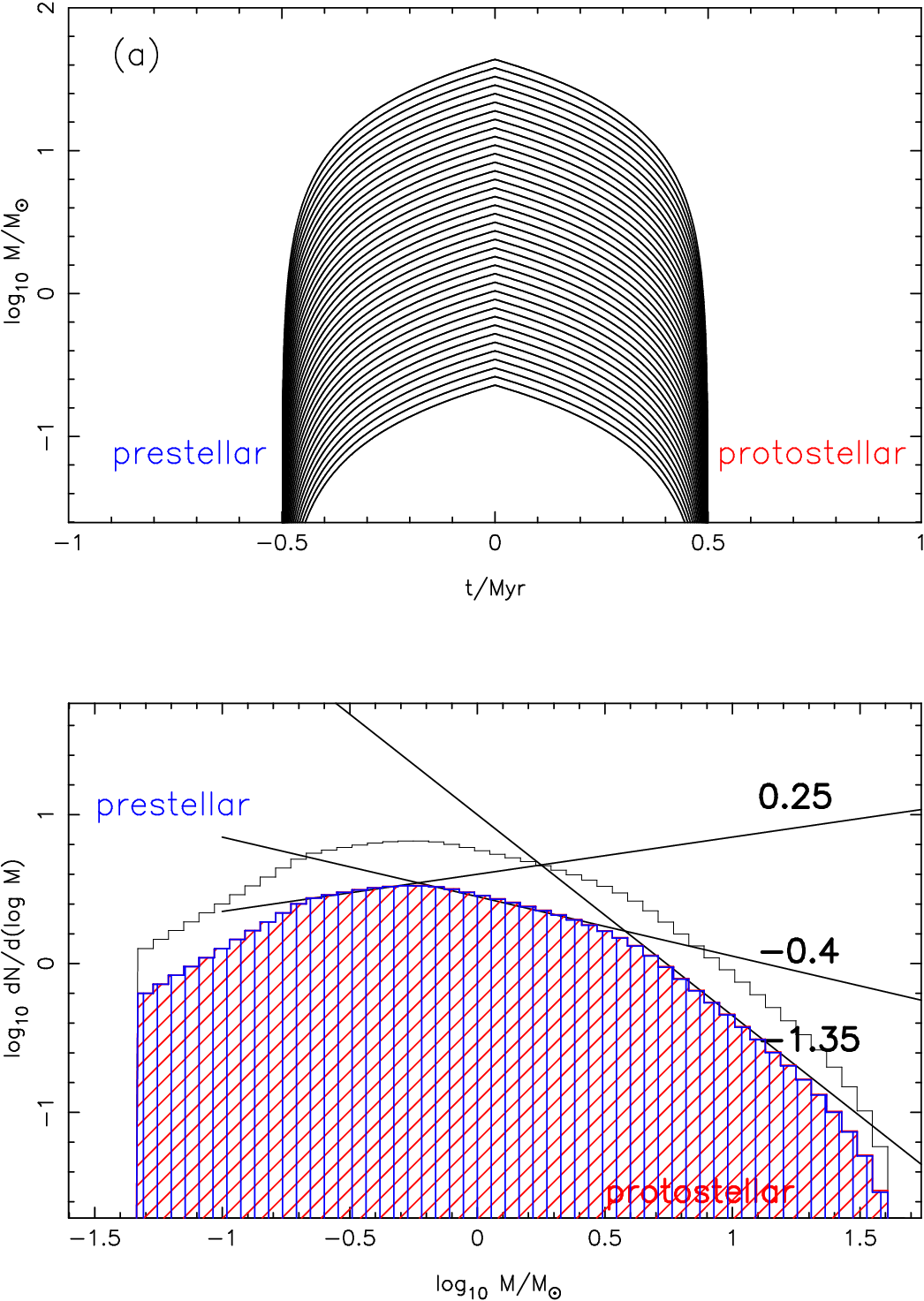}&
\includegraphics[scale=0.3,angle=0]{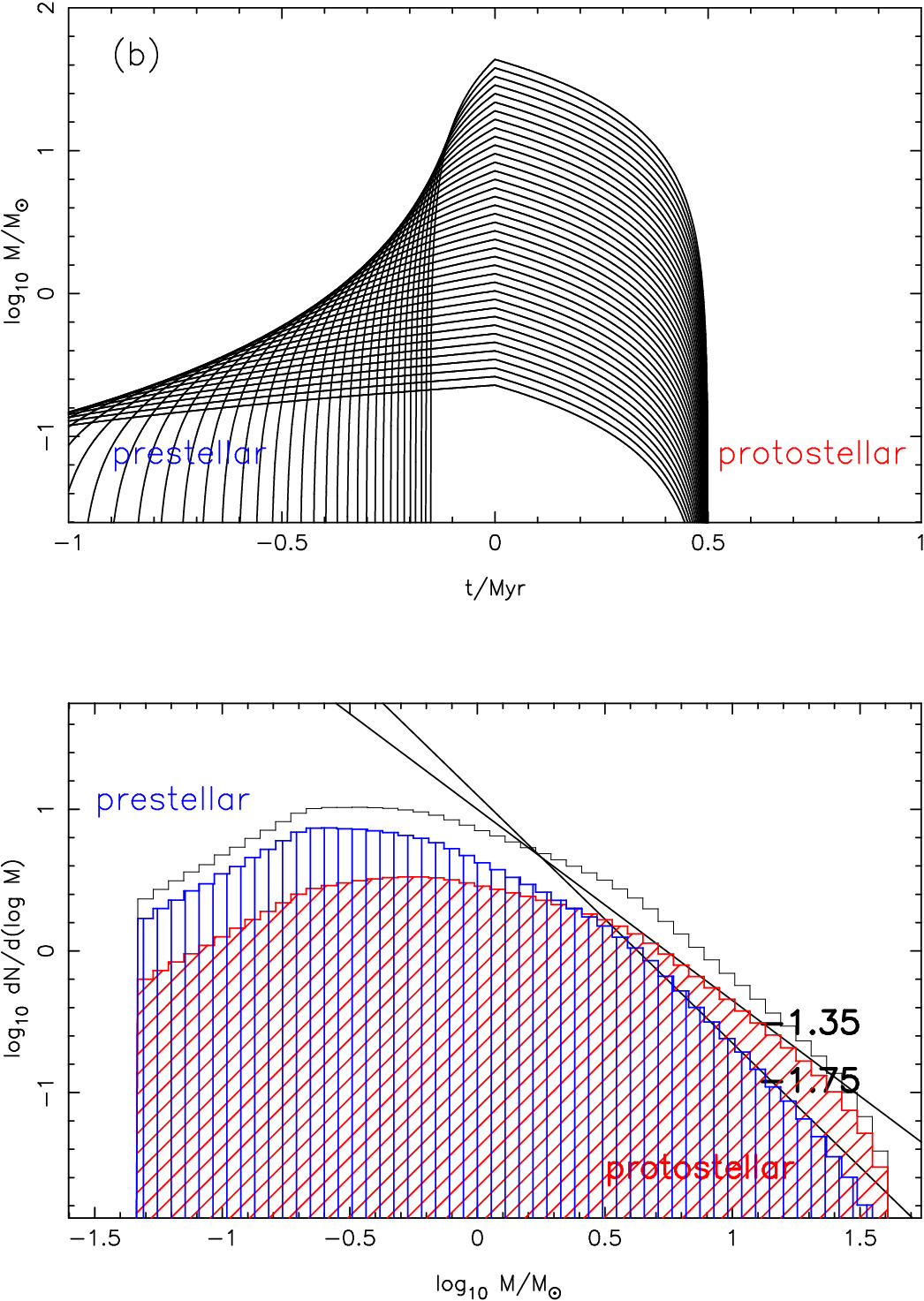}&
\includegraphics[scale=0.3,angle=0]{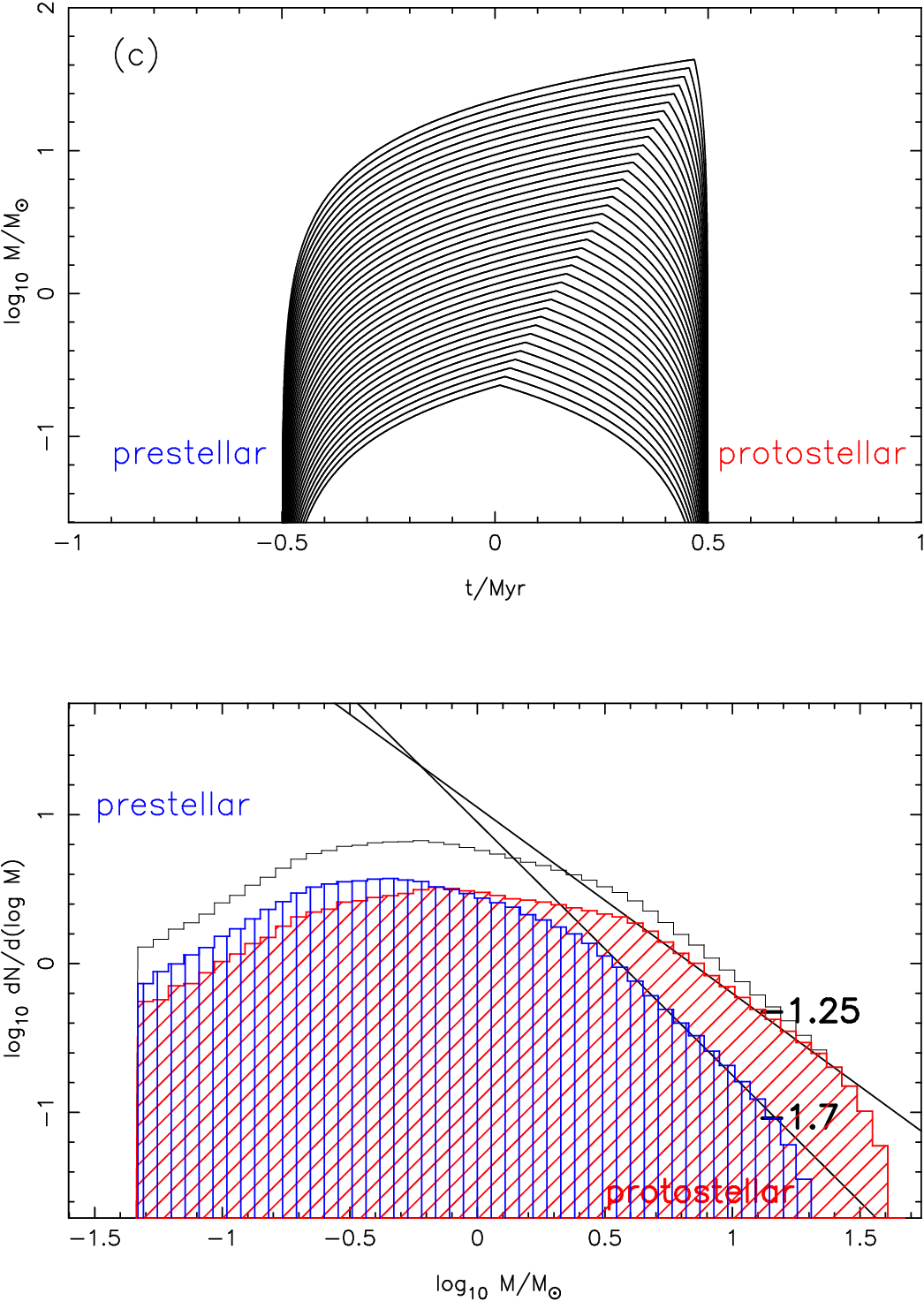}\\
\end{tabular}

\caption{Core mass evolution diagrams (CMEDs) for different
  evolutionary scenarios.  Each top panel shows the evolution of core
  masses as measured above a detection threshold as a function of time
  for various peak masses $m_\mathrm{peak}$, on a linear-log plot.  The bottom
  panels show the corresponding time-averaged core mass functions for
  prestellar cores (blue, vertical hatching), protostellar
  cores (red, diagonal hatching), and all cores (thin
  solid line).  The straight lines are marked with approximate power laws for sections of the CMF.  {\bf (a)} Equal pre/protostellar timescales
  for all cores regardless of mass, with a constant mass
  accumulation/removal rate $\pm\dot m \propto m_\mathrm{peak}$.  {\bf
    (b)} Faster evolution for more massive sources in the
  prestellar phase only.  {\bf (c)} Continuing accretion into the
  protostellar phase, so that the peak mass is reached some time after
  $t=0$.  }
\label{fig:simmass}
\end{figure*}

To represent this graphically we can plot Core Mass Evolution Diagrams
(CMEDs, Figs.~\ref{fig:sim_single} and \ref{fig:simmass}) which show mass vs. time for various
models of core evolution, and the resulting core mass functions.  For
example, in Fig.~\ref{fig:sim_single}a we plot the evolution of a core with a simple linear evolution of mass with time, given by
\begin{equation}
m(t) = m_\mathrm{peak}(1- |t|/T),
\label{eqn:linear}
\end{equation}
where $m_\mathrm{peak}$ is the maximum mass reached by the core, $t=0$ is the time at which a protostar forms, \begin{jh}and $T=0.5\hbox{ Myr}$ the timescale for core growth or mass loss. \end{jh} This represents the accumulation of mass
above the detection threshold during the prestellar phase, and loss of
mass through accretion and outflow as a protostar.  For this model, the time for core growth and the time for a star to form are equal: $t_\mathrm{sf} = t_\mathrm{gr}$.  If we take a population of such cores and observe them at random times then we generate the CMD shown in the lower panel.  The CMD is a power law with a positive index of $+1$, reflecting the longer times
spent in the higher mass bins (which are equal width in $\log_{10}(m)$).  

Likewise, Fig.~\ref{fig:sim_single}b shows cores with evolution given by
\begin{equation}
 m(t) = m_\mathrm{peak}(|t|/T + 1)^{-1/\gamma}
\label{eqn:salpeter}
\end{equation}
with $\gamma = 1.35$ and $T = 10^4 \hbox{ yr}$.  The resulting CMF is a power-law with negative index $-1.35$.  This function, with longer timescales at lower masses, is the pathological
case where time evolution alone reproduces a Salpeter-like CMF from a single population of identical sources.  This mass evolution function becomes unrealistic at low masses because the timescales tend towards infinite (note the apparent sharp cutoff at $\log_{10} (M/\Msun) = -0.85$ is simply the mass reached at the time limits of our calculation).  Nonetheless this, or a similar form for the mass evolution of cores, is interesting as it produces a CMF with a power-law slope at the high mass end similar to the slope of the IMF.  Also, it implies a protostellar mass loss rate which peaks early in the protostellar stage and then decreases with time in line with decreasing outflow mass loss (\citet{batc96}, but see also \citetalias{outflows}).

In such a model, all stars would form inside high mass cores, with low mass stars produced in multiples or with low efficiency.   The strongest evidence against such a scenario is the existence of low-mass low-luminosity Class~0 sources, some of which have small outflows supporting their youth \citep{outflows,dunham07}.  A single peak core mass model is clearly extreme.  However, a population of low-mass cores on the CMD in itself does not imply that protostars form in low mass cores.

To generate more realistic diagrams we assume a population of cores with a distribution of peak (maximum) masses $m_\mathrm{peak}$  following a three-part power-law (the peak core mass function or PCMF):
\begin{equation}
dN(m_\mathrm{peak})/d\log(m_\mathrm{peak})\, \propto m_\mathrm{peak}^{-(\alpha-1)}
\end{equation}
with
\begin{eqnarray*}
\alpha = 0.3 \quad &0.2 < m_\mathrm{peak} \leq 0.8\quad &(-0.7 < \log_{10} m_\mathrm{peak} \leq -0.1)\\\nonumber
\alpha = 1.3 \quad &0.8 < m_\mathrm{peak} \leq 5.0\quad &(-0.1 < \log_{10} m_\mathrm{peak} \leq 0.7)\\\nonumber
\alpha = 2.3 \quad &5.0 < m_\mathrm{peak} \quad&(0.7 < \log_{10} m_\mathrm{peak}). \\\nonumber
\end{eqnarray*}

The boundaries between the power law segments are set a factor of 10 higher in mass than those for the Galactic field IMF \citep{kroupa02} to take into account the star formation efficiency \citep{nutter07,alves07}.  The peak core mass function is significant because it gives the total mass available to each protostar for accretion, which relates directly to the IMF through the SFE (see Sect.~\ref{sect:discussion}).

We then generate CMFs based on this distribution under
  various assumptions about the evolution of the core mass with time.
  From the core mass evolution we calculate the core mass function
  assuming that the ensemble of cores is observed at random points
  during their evolution.  The observed population on the CMF in a
  given mass range is a sum over the population of cores of the time
  spent in that mass range.

In the
simplest case (Fig.~\ref{fig:simmass}a), we assume that the
measured mass above the detection threshold increases at a constant
rate in the prestellar phase, and decreases at a constant rate in the
protostellar phase, following Equation\ref{eqn:linear}.
The timescales for core growth and protostellar formation are equal, $t_\mathrm{gr} = t_\mathrm{sf} = T$.   For this simple
case the observed CMF is similar to the underlying peak mass function
  in that it shows a
  3-part power law above 0.2~\Msun.  However, the power-law indices of $0.25, -0.4$ and $-1.35$ are steeper than those of the underlying peak
  core mass function (0.7,-0.3,-1.3 in $\log m_\mathrm{peak}$) because
  of the contribution of evolving cores detected at less than their maximum
  mass.  This contribution increases towards lower masses, particularly where the underlying mass function has turned over, and also
  explains the population below $\log_{10} m = -0.7$ with index $+1$ which does not
  exist in the PCMF.

Using CMEDs we can now investigate the effects of different
evolutionary scenarios on the observed CMF.

\subsection{Fast evolution of massive cores} 
\label{sect:lifetimes}

 The lack of massive starless cores, and the
 decrease in the number of starless cores towards higher masses, could
 imply that more massive cores have relatively short prestellar
 lifetimes, and that this is why we see few of them.  (The
 alternative, that higher-mass protostars have longer lifetimes, can
 not explain why so few massive starless cores are detected).  Current
 best estimates for the mean lifetime for starless cores detectable in
 the submm are similar to the protostellar lifetime, 1.5--$4 \times
 10^5$~years \citep{paperI,enoch07,jkirk07}, but the free-fall timescales for
 mean densities of $>3\times 10^5$~cm$^{-3}$ are less than $7\times
 10^4$ years, so theoretically this observed phase could be very
 short.

From the observations we can estimate the prestellar to protostellar ratio for cores at
 masses above $10\Msun$: there are four times as many protostars
 suggesting a prestellar lifetime 1/4 of that in the protostellar
 phase.  For comparison, at 3~\Msun\ the number of cores in the
 prestellar and protostellar phases are roughly equal.

 In Fig.~\ref{fig:simmass}b we plot a CMED in which massive cores
 evolve fast in the prestellar phase, with mass evolving
 \begin{jh}linearly with time according to Equation~\ref{eqn:linear} but with varying prestellar timescale 
   $T = 1/\sqrt{m_\mathrm{peak}/\msun}\hbox{ Myr}$.  The protostellar timescale remains constant at $T=0.5\hbox{ Myr}$.\end{jh}  The resulting
   prestellar mass distribution is steeper than the core peak mass
   function (steeper than $m^{-1.35}$) and steeper than the
   (unchanged) protostellar mass distribution.
 Therefore we can reproduce the excess of high-mass protostars by
 assuming protostellar lifetimes are all equal but the prestellar
 lifetimes depend strongly on mass, with short lifetimes for high mass
 sources.  Note that this model still forms stars at the point at which the core stops growing, $t_\mathrm{sf} = t_\mathrm{gr}$.

Longer prestellar lifetimes for lower-mass cores are predicted by
ambipolar diffusion models \citep{tassismouschovias04}, and our
results are qualitatively consistent with this.  Starting from the
assumption that higher mass cores 
evolve from lower density (hence have a higher Jeans
mass), \citet{clark07} find the opposite: that the free-fall timescale
is longer for higher mass cores, and therefore the prestellar mass
distribution should be flatter than the IMF.  Our result differs from
that presented by \citet{clark07} because our cores evolve in mass
during the prestellar and protostellar phase, so there are differences
in how the observed CMF is derived.  Nonetheless, longer timescales
for higher mass clumps are inconsistent with our results.

The above discussion assumes that cores evolve at fixed (gas+stellar)
mass, ie. the timescales for star formation are much less than the
timescales for acquisition of additional material by the cores,
$t_\mathrm{sf} \ll t_\mathrm{gr}$. In this case the difference in mass
distributions argues that massive cores must themselves form and then
evolve to form stars more rapidly than low mass cores.




\subsection{Continuing accretion} 
\label{sect:accretion}

In a model where the core is not considered to be a fixed mass
reservoir but continuing accumulation of mass is allowed during the
protostellar phase, the envelope mass can continue to rise during the
main protostellar accretion phase.  This is the case where the star
formation timescale is less than the timescale for core growth,
$t_\mathrm{sf} < t_\mathrm{gr}$.  The peak envelope mass can then be
reached sometime during the \begin{jh}protostellar\end{jh} phase rather than at the
prestellar/\begin{jh}protostellar\end{jh} transition.   Such models require surrounding
lower-density material to become gravitationally bound to the
core and continue to flow onto the envelopes of existing protostars.

A model of this is shown in the CMED in Fig.~\ref{fig:simmass}c.  In
this case, the time at which the peak core mass is reached is shifted
into the \begin{jh}protostellar\end{jh} phase by an amount
proportional to $\log_{10} m_\mathrm{peak}-\log_{10} m_\mathrm{min}$
where $m_\mathrm{min}= 0.2\Msun$ is the minimum mass in the peak core
mass function, so $t_\mathrm{gr} \geq t_\mathrm{sf}$.  The resulting
prestellar mass distribution is steeper than the peak mass function
and steeper than the combined (pre+protostellar) CMF.  The
protostellar mass function is flatter, reflecting the reduced amount
of time spent at low masses.  Again, this model reproduces the main
features of the observed pre/protostellar mass distributions.  Core
masses which continue to rise into the \begin{jh}protostellar\end{jh} phase can explain why
protostellar cores tend to have masses higher than those of starless
cores.

\begin{jh} Core masses will continue to increase into the protostellar
  phase as long as the mass loss through accretion and outflows is
  outweighed by mass moving from low density at large radii to within
  the column density detection threshold of the observations.
  Although the model in our CMED is clearly simplistic, there are
  several ways in which this kind of behaviour can arise. Recent
  models of core growth through ambipolar diffusion show significant
  inward velocities in the outer parts of the core which continue
  after the central protostar has formed \citep{adamsshu07}.
  Alternatively, if large-scale turbulent flows supply the cores
  \citep[see][]{maclow04rev} then cores will continue to grow until
  the supply of material is exhausted.  Additionally, in competitive
  accretion models, cores which have already accumulated higher masses
  form deeper gravitational wells and competitively accrete gas from
  the surrounding cloud and lower mass cores, which might bias the
  growth towards already massive protostars \citep[][ but note that there are kinematic arguments against competitive
  accretion: \citealt{walsh04,andre07,jorgensen07,kirk07}, though see
  also \citealt{ayliffe07}] {bonnell01}.  \end{jh}


\subsection{Luminosity constraints on mass evolution}

The bolometric luminosity is plotted against core mass for the Perseus
sources in Paper II, Fig. 4 (protostars) and Paper III, Fig. 2 (all
sources). There is a general trend of increasing luminosity with
increasing mass, but with a large scatter. There are sources with an
order of magnitude difference in core mass (the complete sample spans
0.5-50 \Msun) that produce similar bolometric luminosities while there is
an order of magnitude scatter in source luminosity over some ranges of
core mass. In principle the observed luminosity-core mass distribution
provides a constraint on the evolution of the core mass and hence the
CMF, but exploring this is complex.

The mass evolution of cores affects the luminosity of the protostars
forming in the cores through accretion of core material onto the
central star-disk system.  The accretion luminosity is given by
$L_\mathrm{acc} = G M_* \dot M_\mathrm{acc} / R_\mathrm{acc}$ where
$M_*$ is the stellar mass, and $\dot M_\mathrm{acc}$ and
$R_\mathrm{acc}$ are the accretion rate onto the star/disk and the
accretion radius respectively.  However to predict the luminosity
evolution of the sources requires an understanding of how the
accretion rate onto the star/disk $\dot{M}_\mathrm{acc}$ and the rate
of change of mass of the core $\dot{m}$ are related.  This
relationship is likely to be both complex and evolving.  In addition
to the mass loss due to accretion on to the central source, $\dot{m}$
encompasses all the other core mass loss, and growth, mechanisms. This
includes, for example, the mass loss associated with the effect of
winds and outflows (not necessarily self-generated) as well as the
growth of the core due to the accretion of surrounding cloud material.
So for example, in the case where a core continues to grow in mass
once a protostar has formed, $\dot{M}_\mathrm{acc}$ and $\dot{m}$ have
opposite signs up until the point when the core ceases to grow in
mass.  Even in the simple case where $\dot{M}_\mathrm{acc}$ is a
constant fraction of $\dot{m}$, the radius into which the material
falls, $R_\mathrm{acc}$, and hence $L_\mathrm{acc}$, is likely to
evolve, as, for example, a circumstellar disk forms and grows, and
core material of different angular momentum is accreted.

Using various simplifying assumptions and initial conditions, this
luminosity evolution has been modelled by a number of authors
\citep{smith00,froebrich06, myers98,siess97,saraceno96}. However these models have mostly concentrated
on cores which contain a fixed reservoir of material and have not
investigated the evolution of cores as massive the most massive cores
observed in Perseus. It would be interesting to extend such models to
encompass both the range of core masses seen in Perseus and possible
alternative scenarios for the evolution of the core mass, but such
models are beyond the scope of this paper.

\section{From CMF to IMF}
\label{sect:discussion}

So far we have discussed how the observed core mass distributions
relate to the underlying peak core mass distribution, and how
differences between the mass distributions of the prestellar and
protostellar cores might arise.  The subsequent issue is
how these cores go on to form stars with a range of masses consistent
with the IMF. 

\begin{jh}Observational selection effects and systematic errors
  (Section~\ref{sect:assumptions}), as well as small number statistics
  \citep{swift08}, make current measurements of the CMF highly
  uncertain.  Nonetheless, the apparent similarity of the power-law
  slope of the CMF and the IMF has now been noted in several
  regions. \end{jh} The simplest explanation for the match
is that {\em i)} there is a one-to-one mapping between cores and stars
which are formed, {\em ii)} there is a uniform star formation
efficiency (SFE) for all the cores i.e. that the same fraction of the
mass of each core is converted in to star, {\em iii)} the measured
mass of each core is representative of the total mass available to
form a star, and {\em iv)} the evolutionary timescales are
independent of the core mass.

If any of these assumptions is not in fact true, then the mapping of
the CMF to the IMF of the stars formed is clearly more complex.
However we can consider breaking each assumption in turn and examine
how the resulting IMF relates to the CMF as measured for prestellar
cores and/or all submm cores.

\subsection{One-to-one mapping }

Multiplicity breaks the one-to-one mapping between cores and stars.
This has been considered in detail recently by \citet{goodwin07} and \citet{swift08}, who
demonstrate that if multiplicity is significant we would expect the
CMF to be skewed to higher masses compared to the IMF, as massive
cores contribute to a final population of lower-mass stars.
\citet{goodwin07} argue that the best match between the observed CMF
\citep{nutter07} and the IMF \citep{kroupa02} occurs if all cores
(including low-mass cores) form binaries or higher-order multiples.

There is plenty of evidence for multiplicity in Perseus sources.
Although multiple Spitzer MIPS sources can only be identified in 3
cores \citep{jorgensen06a}, at least 10\% of the SCUBA cores have been
demonstrated to contain multiple sources when observed with
high-resolution infrared, radio or mm/submm interferometry
\citep{lay95,anglada00,rac97,rr98,wolfchase00,olinger06,hirano99}, and
the majority of the submm sources have not been the targets of detailed
studies of their fragmentation and multiplicity.

A population of cores which will never form stars, transient cores,
would also break the one-to-one mapping between observed cores and
stars.  Transient cores are more likely to appear at low mass, where
the gravitational binding is weaker, and skew the CMF to lower masses
than the IMF.  The presence of a population of low-mass, transient
cores cannot be ruled out by our data.

\subsection{Star formation efficiency}

If the star formation efficiency (SFE) is not independent of core mass, variations in it can have a complex effect on the relationship between the CMF and IMF.  Changes in the SFE can lead to an IMF which is either steeper or flatter than the CMF or affect different mass ranges differently, depending on exactly how the SFE varies with core mass \citep{swift08}, if indeed the SFE is an unique function of core mass at all.
SFE variation therefore acts to skew or blur the relationship between the CMF and the IMF.

Although there are estimates of the instantaneous star formation
efficiency in cores in Perseus of 10-15\% \citep{jorgensen07}, we have
no information on the variation in SFE between cores.  Future
constraints on the SFE may come from studying the mass transfer
through accretion and outflows and probing how feedback processes
affect star formation within a core.

\subsection{Representative mass measurements}  

If the core masses measured are not representative of the total mass
available for accretion over the lifetime of the core, then the relationship between the IMF and the
CMF is again more complex. As discussed in Sec. 4.1, the observed mass of a core (or
distribution of masses of a sample of cores) is a function of the peak
mass of the cores and the time evolution of their mass. Depending on
these two factors, different ranges of observed core masses may
contribute to the same range of stellar masses in the IMF.  In general
cores will be observed at masses less than their peak mass, skewing
the CMF to lower masses compared to the IMF.  Only if cores spend a
large fraction of their evolution with masses at, or near, their peak
mass is the observed mass of core representative of the total mass
available to form stars, leading to a one-to-one relationship between
the IMF and CMF.

\subsection{Evolutionary timescales}


Longlasting evolutionary phases will be relatively overrepresented in
the count of cores on the CMF.  Conversely, the most rapid phases are
the least likely to be seen.  Therefore, if cores of different peak
masses evolve at different rates, the shape of the CMF is affected.
This is in contrast to the IMF, which takes account of the short
lifetimes of massive stars by means of stellar evolution theory.

The evidence presented here for the absence of massive prestellar cores in
the CMF can be explained if cores with high peak masses evolve more
rapidly.  Therefore, from the observed deficit of massive prestellar
cores we would expect the IMF to be flatter (higher populations at
higher masses) than the CMF.  

\bigskip 
Putting all four factors
together, what can we say about the relationship between the measured
CMF and the IMF?  It is clear that if high-mass prestellar cores are
underrepresented on the CMF, and cores map onto the IMF one-to-one
with a constant core star formation efficiency, then we would expect
the measured CMF to be steeper than the IMF at the high-mass
end.  At the low mass end, the CMF always contains an additional
population of cores evolving to or from higher masses and should
always be overpopulated relative to the IMF.  To reconcile the
observed similarity between the CMF and the IMF, it would be necessary
to invoke two opposing effects, eg. faster evolution for more massive
cores combined with multiplicity, which contrive to cancel out so that
the observed CMF once again, and coincidentally, has the same
power-law form as the IMF.

\section{Summary and conclusions}
\label{sect:summary}


The mass distributions of the prestellar and protostellar cores in
Perseus are different, with an excess of protostellar cores at high
masses and an excess of starless cores at low masses.  There are a
number of possible selection effects which could be influencing the
observed distributions.  Of these, only confusion in regions where
cores and protostars are highly clustered can plausibly explain the
observed distributions.



In clustered regions there is a difficulty in identifying the structures which will ultimately form a star.  This may be because the gas which will go on to form future stars lies in dense,
filamentary structure with no clear boundaries segregating it into discrete, centrally peaked cores.  Confusion in the observations then leads
to both an undercounting of starless cores and also an overattribution
of mass to protostars if they are confused with essentially prestellar
material.  

Of the other assumptions and selection effects involved in deriving
the mass distribution, differences in the temperatures between classes
of cores cannot explain the differences between the mass
distributions.  At low masses, the distributions are affected by
selection effects, particularly our increasing inability to identify
low-luminosity sources as protostellar.  Transient cores may also
contribute to the excess of starless cores at low masses, but seem
unlikely to be significantly influencing the relative underabundance
of massive starless cores.

If selection effects are not the most significant cause of the
difference in mass distributions, then these differences must be
direct consequence of the evolution of cores.  By looking at the mass
distributions of the prestellar and protostellar cores separately, we
can place constraints on core mass evolution. We find two evolutionary
scenarios which explain the observed excess of protostellar cores with
high masses.  The simplest is that timescales for the formation of
protostellar cores may vary with core mass, with more massive cores
forming protostars more quickly than lower mass cores.  Alternatively
(or additionally), the masses above our column density threshold may
continue to rise well into the protostellar phase, rather than peaking
at the point when a protostar is formed.  The observed distributions
rule out the possibility that higher mass cores evolve more slowly to
form protostars than lower mass cores.


Whatever the explanation for the observed differences in the mass
  distributions, they provide a clue to how cores evolve to form
  stars, and how the measured CMF relates to the stellar IMF.  The
relationship between the CMF and the IMF is in general complex and
several assumptions about evolution, detectability and efficiencies have
to be made to derive the IMF from the CMF.  But clearly if we are
undercounting the number of massive starless cores which form, either
because they only last a short time or because of confusion with
protostars, then the IMF should contain more systems at the high mass
end of the distribution than the prestellar CMF.  The IMF should also show relatively fewer sources at the low mass end compared to the CMF, because the CMF contains a contribution from cores observed while evolving to or from their maximum mass.  On the other hand,
we know that many cores will fragment into
multiple lower-mass systems on the IMF \citep[see][]{goodwin07}.  In some clouds
  these effects may conspire to cancel each other, producing a CMF and
  an IMF with similar shapes.

Clouds such as Perseus where the distribution of prestellar and
  protostellar core masses are inconsistent with an apparently simple,
  direct mapping of the CMF to IMF provide important insights into the
  possible evolution of cores as they form stars.  Deeper extensive
  surveys of cores and protostars in other clouds are needed to help
  unravel the details of these possible evolutionary paths of the
  cores.

\acknowledgements

The authors would like to thank Nicolas Peretto and Daniel Price for
useful discussions, and referee Doug Johnstone for suggestions which significantly improved the paper. The James Clerk Maxwell Telescope is operated by
the Joint Astronomy Centre on behalf of the Science and Technology
Facilities Council of the United Kingdom, the Netherlands Organisation
for Scientific Research, and the National Research Council of Canada.
This work is based in part on observations made with the Spitzer Space
Telescope, which is operated by the Jet Propulsion Laboratory,
California Institute of Technology under a contract with NASA.  JH
acknowledges support from the PPARC (now STFC) Advanced Fellowship programme.

\bibliographystyle{aa}
\bibliography{perseus}

\end{document}